\documentstyle[aps,preprint]{revtex}
\begin{document}
\draft
\title{\Large \bf The Thermodynamics of the XXZ Heisenberg Chain with
                       Impurities}
 \author{Boyu Hou$^a$, Kangjie Shi$^{a,b}$, Ruihong  Yue$^{a,c}$, Shaoyou 
                  Zhao$^a$}
 \address{ \footnotesize $^a$ Institute of modern physics, Northwest
university,P.O.Box 105,\\ Xi'an 710069, P.R. China\\
  $^b$ CCAST (World Laboratory), P.O. Box 8730, Beijing
100080, China\\
  $^c$ Institute of  Theoretical Physics, Academia Sinica, 
Beijing 100080}
\date{Sept, 1 1999} 
\maketitle
\narrowtext
\begin{abstract}
  In this paper, we discuss the effect of the arbitrary spin impurities to
the spin-1/2 and spin-1 XXZ model. The effect of ground state, the free
energy, the magnetic susceptibility, the specific heat and the Kondo
temperature are  given by using
the thermodynamic Bethe ansatz equation.    
\end{abstract}  
\narrowtext
\section{Introduction}
\par Behavior of a quantum chain  will be quite different when impurities
embedding in the system. With the method of bosonization and
renormalization techniques, Kane and Fisher proved this result when they
studied the properties of a potential scattering center in a Luttinger
liquid \cite{KF}. After their work, many papers have paid their attention
to the
problems of local perturbation to a Luttinger liquid, especially the 
Kondo problem in the system [2-5].  It is well known that the spin
dynamics of
the Kondo problem is equivalent  to the  dynamics of the spin chain with
magnetic impurities \cite{NK}. 

   The Heisenberg chain is an important model in the study of integrable
system. Firstly, Andrei and Johannesson considered a spin-s 
($s>\frac{1}{2}$) impurity embedding in the spin-$\frac{1}{2}$ Heisenberg
chain with periodic boundary condition \cite{AJ}.  Then, Lee and Schlottmann
generalized the problem to arbitrary spin \cite{LS}.  However, it is found
that the open boundary theory is better to deal with quantum chains with
impurities, since the
periodic boundary condition would lead to the presence of some unphysical
terms in the Hamiltonian to maintain the integrability \cite{W,HP}. In
resent research, in the frame of the quantum inverse scattering method ,
Wang studied the exact solution of the spin-$\frac{1}{2}$ XXX open
Heisenberg chain with
two arbitrary spin impurities \cite{W}. Subsequently, Hu and Pu considered
the
spin-$\frac{1}{2}$ XXZ chain with two spin-$\frac{1}{2}$  impurities
\cite{HP}. In the previous  paper \cite{zsy}, we obtained the eigenvalues
of the Hamiltonian and the Bethe ansatz equations for both the
spin-$\frac{1}{2}$ and spin-1 XXZ chains with two
arbitrary spin impurities.      

  The low-temperature thermodynamics was first proposed by Yang and Yang. 
In their series papers, they studied systematically the ground state, the
magnetic susceptibility and the specific heat of the spin-$\frac{1}{2}$
Heisenberg chain \cite{y1,y2,y3}. Directly, this method was generalized to
other
models,
such as the Hubbard model {\sl etc} \cite{tak,gau,john}. In the Ref.
\cite{bab,resh,mez,pasq,alca},
Babujian, Kirillov and Reshetikhinthe and others studied thermodynamics of
the arbitrary spin XXX and XXZ Heisenberg chain respectively.  In the
present paper, we study thermal properties of the spin-$\frac{1}{2}$ and
spin-1 XXZ chains with two arbitrary spin impurities by using the standard
method at low temperature. 

\section{The model}
\par
The integrability of the spin-$\frac{1}{2}$ and spin-1
XXZ Heisenberg chains with two arbitrary spin impurities have been
discussed in the Ref.{\cite{zsy}}. Hamiltonians of them can be 
represented by
\begin{equation}
H=\left.
\frac{-i}{2\rho^{N}(0)\rho_d(c_a)\rho_d(c_b)tr_\tau
    K^+(0)}\times\frac{d\;t(u)}{du}\right
|_{u=0},
\end{equation}
where $t(u)$ is the transfer matrix,
$\rho(0)$ and $\rho_d(c_i)\;(i=a,\;b)$ are constants appearing in  
unitarity relations of the $R$-matrix in the bulk and the
boundary respectively, $K^+$ is the dual
reflection matrix, and $c_i$ are coupling constants.

Applying the Hamiltonians $H$ on the multi-particle-state Bethe vector 
$|\Omega\rangle $, we obtain the eigenvalues as follows
\begin{equation}
E=E_b+E_i+E_\infty \; ,
\end{equation}
where $E_b$ , $E_i$ and $E_\infty$ stand for the energy of boundary ,
impurities and the bulk respectively. To the spin-$\frac{1}{2}$ chain ,
we have 
\begin{eqnarray}
&&E_b=-\frac{i\sinh(\xi^+-\xi)}{2\sinh\xi^+\sinh\xi}
                  +\frac{i}{2\sinh(2\eta)} 
\; ,\nonumber\\[3mm]
&&E_i=-\frac{i}{2}\sum_{r=\pm 1}\sum_{k=a,b}
        \coth(rc_k+s_{i}\eta+\frac{\eta}{2})\;
                      ,\nonumber\\[3mm]
&&E_\infty=-N\frac{i\cosh\eta}{\sinh\eta}
     -\sum^M_{j=1}\frac{2i\sinh\eta}{\cosh(2v_j)-\cosh\eta} \; ,     
\end{eqnarray}
where $s_{i}$ is the arbitrary spin of impurities; to spin-1 chain case,
we
have
\begin{eqnarray}
E_b&=&-\frac{i\sinh(2\eta)}
           {2\sinh(\xi^++\frac{\eta}{2})\sinh(\xi^+-\frac{3\eta}{2})}
                +2\frac{i\cosh\eta}{\sinh(3\eta)} \;
,\nonumber\\[5mm]
E_i&=&-\sum_{r=\pm
1}\sum_{k=a,b}\frac{i\sinh(2rc_k+2s_{i}\eta+\eta)}
               {2\sinh(rc_k+s_{i}\eta)\sinh(rc_k+s_{i}\eta+\eta)}\;
,                 \nonumber\\[5mm]
E_\infty&=&-N\frac{i\sinh(3\eta)}{\sinh\eta\sinh(2\eta)}
         -2N\frac{i\cosh(2\eta)}{\sinh(2\eta)}
 -\sum_{j=1}^M\frac{2i\sinh(2\eta)}{\cosh(2v_j)-\cosh(2\eta)},
\end{eqnarray} 
where $\xi^+,\;\xi $ and $\eta$ are some free parameters and $v_j$
satisfies the Bethe ansatz equation
\widetext
\begin{eqnarray}
& &\frac{\sinh(v_j-\xi^+-\frac{\eta}{2})\sinh(v_j+\xi-\frac{\eta}{2})
\sinh(v_j+\frac{\eta}{2})\cosh(v_j+\frac{\eta}{2})\sinh^{2N}(v_j+s\eta)}
{\sinh(v_j+\xi^++\frac{\eta}{2})\sinh(v_j-\xi+\frac{\eta}{2})
\sinh(v_j-\frac{\eta}{2})\cosh(v_j-\frac{\eta}{2})\sinh^{2N}(v_j-s\eta)}
                   \nonumber\\
& &\times   \prod_{r=\pm 1}\prod_{k=a,b}
               \frac{\sinh(v_j+rc_k+s_{i}\eta)}
                    {\sinh(v_j+rc_k-s_{i}\eta)}
=-\prod_{i=1}^M\frac{\sinh(v_j-v_i+\eta)\sinh(v_j+v_i+\eta)}
                          {\sinh(v_j-v_i-\eta)\sinh(v_j+v_i-\eta)} \;,
\end{eqnarray}
\narrowtext
where $j=1,\; 2,\; \cdots,\; M$ with $M$ being the number of down spins.
 
With the Bethe vector $|\Omega\rangle$, according the Ref. \cite{bab,resh}, 
we can also obtain the
eigenvalue of the total spin $z$-projection
\begin{equation}
s^z=Ns+2s_{i}-M
\end{equation}

\section{The Ground state}
\par  To obtain precise forms of the 
eigenvalues of Hamiltonians, denoting 
\begin{equation}
\Phi_\gamma(v_j)=2\tan^{-1}\left [ \cot\frac{\gamma}{2}\tanh v_j\right ],
\end{equation}
and taking logarithm of the Bethe ansatz equation (5), we have 
 \begin{eqnarray}
& &-\Phi_{2\xi^++\gamma}(v_j)+\Phi_{2\xi-\gamma}(v_j)+\Phi_\gamma(v_j)
-\Phi_{\pi+\gamma}(v_j) +2N\Phi_{2s\gamma}(v_j)\nonumber \\
&+&\sum_{k=a,b}\sum_{r=\pm
1}\Phi_{2S_{i}\gamma}(v_j+rc_k) = 
          2\pi I_j+\sum_{i=1}^M\left\{\Phi_\gamma(v_j-v_i)
                         +\Phi_\gamma(v_j+v_i)\right \},
\end{eqnarray}   
where the transformations $\eta\rightarrow i\gamma,\;\; \xi 
\rightarrow i\xi$ and $\xi^+\rightarrow i\xi^+$ have been used,  and $I_j$
is an integer. Taking the thermodynamic limit ( $N\rightarrow\infty$ ),
the distribution function of particles can be
considered as continuous one. So the Bethe ansatz
equation can be written as
\begin{equation}
\rho(v)=\frac{1}{2\pi}\left
[\Phi'_{2s\gamma}(v)+\frac{1}{2N}\theta'(v)\right]
   -\frac{1}{2\pi}\int_{-\infty}^{\infty}d\nu
\Phi'_{2\gamma}(v-\nu)\rho(\nu)
\end{equation}
where the distribution function is defined by\cite{y1}
\begin{equation}
\rho(v)=\frac{dI_j(v)}{dv},
\end{equation}
and $$ \theta(v)\equiv
-\Phi_{2\xi^++\gamma}(v)+\Phi_{2\xi-\gamma}(v)+\Phi_\gamma(v)
+\Phi_{\pi+\gamma}(v_j)+\sum_{k=a,b}\sum_{r=\pm
1}\Phi_{2s_{i}\gamma}(v+rc_k).$$   
 Here the $\theta(v)$ stands for the effects of open boundary
condition and  magnetic impurities. Applying the Fourier transformation on
eq.(9), we have
\begin{eqnarray}
\hat
\rho(\omega)&=&\frac{\sinh(\pi\omega-2s\gamma\omega)}
                    {2\sinh(\pi\omega-\gamma\omega)\cosh(\gamma\omega)}
+\frac{1}{2N}\left\{
\frac{\cosh(\frac{\pi}{2}\omega-\gamma\omega)\sinh(\frac{\pi}{2}\omega)}
        {\sinh(\pi\omega-\gamma\omega)\cosh(\gamma\omega)}
  \right.              \nonumber \\
& &+\;
   \frac{\cosh[(\pi-\xi-\xi^+)\omega]\sinh[(\xi^+-\xi+\gamma)\omega]}
        {\sinh(\pi\omega-\gamma\omega)\cosh(\gamma\omega)}
                          \nonumber \\
& & +\;
\left. B(\omega)\frac{\sinh(\pi\omega-2s_{i}\gamma\omega)}
       {2\sinh(\pi\omega-\gamma\omega)\cosh(\gamma\omega)}\right\},
\end{eqnarray}
where the $B(\omega)=4\cos[c_a+c_b)\omega]
\cos[(c_a-c_b)\omega]$. 

 The distribution function $\rho(v)$ can be obtained by
\begin{equation}
\rho(v)=\int_{-\infty}^{\infty}\hat\rho(\omega)
                     e^{-2iv\omega}d\omega.
\end{equation} 
The ground state energy per site
for our systems can be written as \\

\begin{equation}
\frac{E}{N}=\frac{E_i}{N}+\frac{E_b}{N}+\frac{E_\infty}{N}.
\end{equation}
(i). when $s=\frac{1}{2}$, after some calculation, we have the ground
state energy of the bulk term
\begin{equation}
E_\infty=
 - N\frac{\cosh\gamma}{\sinh\gamma}+N\int_{-\infty}^\infty
           dv\frac{\sin(\gamma)}{\cosh(2s\gamma)-\cos(2v)}
\end{equation}
It is exactly same as one in the periodic case \cite{y1}. The
contributions 
of the boundary ( $E_b$ ) and the impurities ( $E_i$ ) are
\begin{eqnarray}
E_b&=&-\frac{\sinh(\xi^+-\xi)}
   {2\sinh\xi^+\sinh\xi} +\frac{1}{2\sinh(2\eta)}\nonumber\\
& & -\int_{-\infty}^{\infty}d\omega
   \frac{\sinh[(\xi^+-\xi+\gamma)\omega]\cosh[(\pi-\xi^+-\xi)\omega]}
        {\sinh(\pi\omega)\cosh(\gamma\omega)}\nonumber\\
& & -\int_{-\infty}^{\infty}d\omega
    \frac{\cosh(\pi\omega/2-\gamma\omega)\sinh(\pi\omega/2)}
         {\sinh(\pi\omega)\cosh(\gamma\omega)}\\
E_i&=&\frac{i}{2}\sum_{r=\pm 1}\sum_{k=a,b}
            \coth(rc_k+is_{i}\gamma+\frac{i\gamma}{2})\nonumber\\
& &     -\int_{-\infty}^{\infty}d\omega
       \frac{4\sinh[(\pi-2s_{i}\gamma)\omega]\cos[(c_a+c_b)\omega]  
             \cos[(c_a-c_b)\omega]}
              {\sinh(\pi\omega)\cosh(\gamma\omega)};
\end{eqnarray}

(ii). when $s=1$, similarly, we have
\begin{eqnarray}
E_\infty &=&-N\frac{\sinh(3\gamma)}{\sinh\gamma\sinh(2\gamma)}
              -2N\frac{\cosh(2\gamma)}{\sinh(2\gamma)}\nonumber\\
      & &-N\int_{-\infty}^{\infty}d\omega
           \frac{\sinh(\pi\omega-2\gamma\omega)}
                    {2\sinh(\pi\omega)\cosh(\gamma\omega)},\nonumber\\
E_b&=&-\frac{\sinh(2\gamma)}
           {2\sinh(\xi^++\frac{\gamma}{2})\sinh(\xi^+-\frac{3\gamma}{2})}
                +\frac{2\cosh\gamma}{\sinh(3\gamma)}\nonumber\\
& & -\int_{-\infty}^{\infty}d\omega
   \frac{\sinh[(\xi^+-\xi+\gamma)\omega]\cosh[(\pi-\xi^+-\xi)\omega]}
        {\sinh(\pi\omega)\cosh(\gamma\omega)}\nonumber\\
& & -\int_{-\infty}^{\infty}d\omega
    \frac{\cosh(\pi\omega/2-\gamma\omega)\sinh(\pi\omega/2)}
         {\sinh(\pi\omega)\cosh(\gamma\omega)}\\
E_i&=&-\sum_{r=\pm
1}\sum_{k=a,b}\frac{i\sinh(2rc_k+2is_{i}\gamma+i\gamma)}
{2\sinh(rc_k+is_{i}\gamma)\sinh(rc_k+is_{i}\gamma+i\gamma)}\;
                  \nonumber\\ 
& &    -\int_{-\infty}^{\infty}d\omega
       \frac{4\sinh[(\pi-2s_{i}\gamma)\omega]\cos[(c_a+c_b)\omega]
             \cos[(c_a-c_b)\omega]}
              {\sinh(\pi\omega)\cosh(\gamma\omega)}.
\end{eqnarray}          
       
\section{TBA and Physical Properties}
It is well known that that the general solution of Bethe ansatz equation
(5) in the thermodynamic limit lies in the complex plane and forms strings
with the length $n$ \cite{y3,bax}:
\begin{equation}
v_j\equiv \lambda_{\alpha, j}^n=\lambda_{\alpha}^n+i\frac{\gamma}{2}
                (n+1-2j)\;\;\;\;\;\;(\;j=1,\;2,\,\cdots\; ,\; n\;),
\end{equation}
where $n=1,\; 2,\;\cdots,\; \infty$, and $\alpha=1,\; 2,\;\cdots,\;
M^{(n)}$. Here the number of $n$-strings $M^{(n)}$ 
satisfies 
$\sum_{n=1}^\infty nM^{(n)}=M. $
Substituting the above string solution into eq.(5), we have 
\begin{equation}
2N\psi_n^s(\lambda_\alpha^n)+\theta_n(\lambda_\alpha^n)=2\pi I_\alpha^n 
+\sum_{m,\beta>0}\left\{\varphi_{mn}(\lambda_\alpha^n-\lambda_\beta^m)
          +\varphi_{mn}(\lambda_\alpha^n+\lambda_\beta^m)\right\}. 
\end{equation}               
Here,\begin{eqnarray*}
\psi_n^s(\lambda_\alpha^n)&=&\left\{\begin{array}{ll} 
    \Phi_{n\gamma}(\lambda_\alpha^n) & \;\;\;\;\;\;\; s=\frac{1}{2} \\
    \Phi_{\gamma(n+1)}(\lambda_\alpha^n)+
           \Phi_{\gamma(n-1)}(\lambda_\alpha^n) &
          \;\;\;\;\;\;\;s=1 ,
                \end{array}\right.
 \\
 \theta_n(\lambda_\alpha^n)&=& \Phi_{n\gamma}(\lambda_\alpha^n)
          -\Phi_{\pi+n\gamma}(\lambda_\alpha^n)
          -\Phi_{\gamma(n-\frac{2}{\gamma}\xi^+-2j)}(\lambda_\alpha^n)
          +\Phi_{\gamma(n+\frac{2}{\gamma}\xi-2j)}(\lambda_\alpha^n)
                    \\
& &   +\sum_{r,k}\Phi_{\gamma(n+2s_{i}+1-2j)}(\lambda_\alpha^n+rc_k)
                  \\
  \varphi_{mn}(\lambda_\alpha^n\pm\lambda_\beta^m)&=&
(1-\delta_{mn})\Phi_{\gamma|m-n|}(\lambda_\alpha^n\pm\lambda_\beta^m)
       +2\Phi_{\gamma(|m-n|+2)}(\lambda_\alpha^n\pm\lambda_\beta^m)
            \\ & &     +\cdots 
       +2\Phi_{\gamma(|m+n|-2)}(\lambda_\alpha^n\pm\lambda_\beta^m)
       +\Phi_{\gamma(m-n)}(\lambda_\alpha^n\pm\lambda_\beta^m)  
\end{eqnarray*}
Under the thermodynamic limit, according to \cite{y3},
we obtain the following equation
\begin{equation}
\rho_n^h(\lambda)=\frac{1}{2\pi}[(\psi^s_n)'(\lambda)
                    +\frac{1}{2N}\theta_n'(\lambda)]
        -\sum_{m\ge 1}\int_{-\infty}^{\infty}d
                 \mu A_{nm}(\lambda-\mu)\rho_n(\mu) ,
\end{equation}
where $\rho_n$ and $\rho_n^h$ are the distribution functions of the
particle and the hole respectively, and 
$$ A_{nm}(\lambda-\mu)
=\frac{1}{2\pi}[\varphi'_{nm}+2\pi\delta_{nm}\delta(\lambda-\mu)]. $$

The energy of per cite can be written as
\begin{equation}
\frac{E}{N}=\frac{1}{N}E^s_f-\sum_{n\ge
1}\int_{-\infty}^{\infty}d\lambda(\psi^s_n)(\lambda)\rho_n(\lambda),
\end{equation}
where $$
E_f^s=E_b^s+E_i^s+\left\{\begin{array}{ll}
      N\cosh\eta & \;\;\;\;\;\;s=\frac{1}{2} \\
     \displaystyle N\frac{\sinh(3\eta)}\sinh\eta+2N\cosh(2\eta)
&\;\;\;\;\;\;s=1
        \end{array}\right. 
  $$ 
The entropy of per cite can be obtained after we analysis the distributions
of particles and holes. It reads
\begin{equation}
\frac{S}{N}=\sum_{n\ge 1}\int_{-\infty}^{\infty}d\lambda\left\{
(\rho_n(\lambda)+\rho_n^h(\lambda))\ln(\rho_n(\lambda)+\rho_n^h(\lambda))  
 -\rho_n(\lambda)\ln\rho_n(\lambda)-\rho_n^h(\lambda)\ln\rho_n^h(\lambda)
 \right\}.
 \end{equation}
The eigenvalue of the spin z-projection can be obtained with the
so-called Bethe vector. With the distribution function, we have
\begin{equation}
\frac{s^z}{N}=s+\frac{2}{N}s_{i}-\sum_{n\ge 1}n
    \int_{-\infty}^{\infty}\rho_n(\lambda)d\lambda
\end{equation}

In order to find the equilibrium distribution functions at the finite
temperature $T$, we must minimize the free energy
\begin{equation}
F=E-TS-HS^z
\end{equation}
with $\delta F=0$ subject to the constrain eq. (22). Then we obtain the 
thermodynamic Bethe ansatz (TBA) equation
\begin{equation}
\ln(1+\frac{\hat\rho_n^h(\omega)}{\hat\rho_n(\omega)})
 =\frac{1}{T}(Hn-\frac{1}{2}(\psi^s_n)'(\omega))+\sum_{m\ge 1}\hat A_{nm}
\ln(1+\frac{\hat\rho_n(\omega)}{\hat\rho_n^h(\omega)}),  
\end{equation}
where symbols with a hat are results after Fourier transformation on
them, and 
\begin{eqnarray}
\hat
A_{nm}(\omega)&=&\frac{2\coth(\gamma\omega)\sinh[min(m,n)\gamma\omega]
                    \sinh[\pi\omega- max(m,n)\gamma\omega]}
                    {sinh(\pi\omega)}, \\
\hat A^{-1}_{nm}(\omega)&=&\delta_{nm}-\frac{1}{2\cosh(\gamma\omega)}
      \left(\delta_{n.m+1}+\delta_{n.m-1}\right ).
\end{eqnarray}
Define 
\begin{equation}
\hat\varepsilon_n(\omega)=T\ln\frac{\hat\rho_n^h(\omega)}{\hat\rho_n(\omega)}.
\end{equation}
Then, the above TBA equation can be rewritten as 
\begin{equation}
\ln(1+e^{\hat\varepsilon_n(\omega)/T})
       =\frac{1}{T}(Hn-\frac{1}{2}(\psi^s_n)'(\omega))
  +\sum_{m\ge 1}\hat A_{nm}\ln(1+e^{-\hat\varepsilon_n(\omega)/T})
\end{equation}            
Solving the $\hat\varepsilon_n(\omega)$ from the above equation, we 
have
\begin{eqnarray}
\hat\varepsilon_1(\omega)&=&T\hat p
              \ln(1+e^{\hat\varepsilon_{n+1}(\omega)/T}) \\ 
\hat\varepsilon_n(\omega)&=&T\hat p
               \ln[(1+e^{\hat\varepsilon_{n+1}(\omega)/T})
                   (1+e^{\hat\varepsilon_{n-1}(\omega)/T})]
               -\pi \hat p \delta_{n,2s}
\end{eqnarray}
where $\hat p\equiv \frac{1}{2\cosh(\gamma\omega)}$. From eq. (28), it is 
easy to find that
\begin{equation}
\lim_{n\rightarrow\infty}\frac{\varepsilon_n}{n}=H.
\end{equation}
Taking into account (26) and (27), the free energy per site at
equilibrium state is given by
\begin{eqnarray}
\frac{F}{N}&=&\frac{1}{N}E_f-2\tan(2n\gamma)-H(s+\frac{1}{N}s_{i})
\nonumber \\
& &-\frac{T}{2\pi}\sum_{n\ge
1}\int_{-\infty}^\infty d\lambda (\psi^s_n)(\lambda)
      \ln(1+e^{-\varepsilon_n(\lambda)/T})\nonumber \\
& &-\frac{1}{N}\frac{T}{2\pi}\sum_{n\ge
1}\int_{-\infty}^\infty
d\lambda\frac{\theta'_n(\lambda)}{(\psi^s_n)'(\lambda)}(\psi^s_n)'(\lambda)
      \ln(1+e^{-\varepsilon_n(\lambda)/T})
\end{eqnarray}

We can now obtain the specific heat in the condition $T\rightarrow 0,\;
H=0$, and the magnetic susceptibility at the small magnetic field by using
the following formulas 
\begin{eqnarray}
C_s&=&-T\frac{\partial}{\partial T}\left.\left (\frac{S}{N}\right )
             \right |_{H}
     =-T\frac{\partial^2}{\partial T^2}\left.\left (\frac{F}{N}
                        \right )\right |_{H}\\
\chi_s&=&-\frac{\partial^2}{\partial H^2}\left.\left (\frac{F}{N}\right
)\right |_{T}
\end{eqnarray}
Substituting  eq.(34) into eq.(35) and (36), we find that the former three
terms are zero and the fourth term is exactly same as the periodic case
which has been discussed in Ref. \cite{tak,bab,resh,mez}. Then,
we shall discuss the last term of the eq. (34). 
As before, the $\theta_n(\lambda)$ can be divided into the boundary and
the impurity terms 
\begin{eqnarray}
\theta^b_n(\lambda)&=&\Phi_{n\gamma}(\lambda_\alpha^n)
          -\Phi_{\pi+n\gamma}(\lambda_\alpha^n)
          -\Phi_{\gamma(n-\frac{2}{\gamma}\xi^+-2j)}(\lambda_\alpha^n)
          +\Phi_{\gamma(n+\frac{2}{\gamma}\xi-2j)}(\lambda_\alpha^n),\\
\theta^i_n(\lambda)&=&
         \sum_{r,k}\Phi_{\gamma(n+2s_{i}+1-2j)}(\lambda_\alpha^n+rc_k). 
\end{eqnarray}
Then, the last term of eq.(34) can be rewritten as
\begin{eqnarray}
\frac{\tilde F}{N}&=&-\frac{1}{N}\frac{T}{2\pi}\sum_{n\ge
1}\int_{-\infty}^\infty
d\lambda\frac{(\theta^b_n)'(\lambda)}{(\psi^s_n)'(\lambda)}(\psi^s_n)'(\lambda)
      \ln(1+e^{-\varepsilon_n(\lambda)/T}) \nonumber  \\
    & & -\frac{1}{N}\frac{T}{2\pi}\sum_{n\ge
1}\int_{-\infty}^\infty
d\lambda\frac{(\theta^i_n)'(\lambda)}{(\psi^s_n)'(\lambda)}(\psi^s_n)(\lambda)
      \ln(1+e^{-\varepsilon_n(\lambda)/T})\nonumber\\
&=&-\frac{1}{N}\frac{T}{2\pi}\sum_{n\ge
1}\int_{-\infty}^\infty
d\omega\frac{(\hat\theta^b_n)'(\omega)}
            {(\hat\psi^s_n)'(\omega)}(\hat\psi^s_n)'(\omega)
      \ln(1+e^{-\hat\varepsilon_n(\omega)/T})\nonumber \\
& &-\frac{1}{N}\frac{T}{2\pi}\sum_{n\ge
1}\int_{-\infty}^\infty
d\omega\frac{(\hat\theta^i_n)'(\omega)}
            {(\hat\psi^s_n)'(\omega)}(\hat\psi^s_n)'(\omega)
      \ln(1+e^{-\hat\varepsilon_n(\omega)/T})
\end{eqnarray}
In this paper, we mainly discuss the effect of the impurities so here we
only give the result about the last term of the eq.(39).
For both spin-$\frac{1}{2}$ and spin-1 cases, 
\begin{equation}
(\hat\theta^i_n)'(\omega)=\left \{
   \begin{array}{ll}
    \displaystyle   2\pi B(\omega)\frac{\sinh(2s_{i}\gamma\omega)
               \sinh(\pi\omega-n\gamma\omega)}
               {\sinh(\pi\omega)\sinh(\gamma\omega)} & n>2s_{i} \\[5mm]
    \displaystyle   2\pi B(\omega)\frac{\sinh(n\gamma\omega)
                \sinh(\pi\omega-2s_{i}\gamma\omega)}
                  {\sinh(\pi\omega)\sinh(\gamma\omega)} & n\le 2s_{i}  
    \end {array}                       \right.   ,
\end{equation}
where $B(\omega)=4\cos[(c_a+c_b)\omega]
               \cos[(c_a-c_b)\omega]$.  
And 
\begin{eqnarray}
(\hat\psi^s_n)'(\omega)&=&2\pi 
\frac{\sinh(\pi\omega-n\gamma\omega)}{\sinh(\pi\omega)},
\;\;\;\;\;\;\;\;\;\;\;\;\;\;\;\;\;\;\;\;\;\;\;\;\;\;\;\;\;\;
           (\;s=\frac{1}{2}\;) \\
   (\hat\psi^s_n)'(\omega)&=&
        2\pi\frac{2\cosh(\gamma\omega)\sinh(\pi\omega-n\gamma\omega)}
                 {\sinh(\pi\omega)}.
\;\;\;\;\;\;\;\;\;\;\;\;\;\;(\;s=1\;) 
\end{eqnarray}
Substituting the results of $(\hat\theta^i_n)'(\omega)$,
$(\hat\psi^s_n)'(\omega)$ into eq.(39), we can
obtain the effect of the impurities to free energy and other physical
quantities. However, the calculation is much more
complicated than we expected.  Fortunately,  comparing with the periodic
case, we can obtain the low-temperature specific
heat and the magnetic susceptibility contributed by impurities as 

\begin{eqnarray}
C_{1/2}^i&=& \frac{2C_{1/2}^0}{N}\times
    \cosh\left[\frac{(c_a+c_b)\pi}{2\gamma}\right]           
    \cosh\left[\frac{(c_a-c_b)\pi}{2\gamma}\right]
     \frac{\sin\left[\pi s_{i}-\frac{(2K+1)\pi^2}{2\gamma}\right]}
     {\cos(\frac{\pi^2}{2\gamma})}; \\
C_{1}^i&=&\frac{2C_{1}^0}{N}\times
    \cosh\left[\frac{(c_a+c_b)\pi}{2\gamma}\right]
    \cosh\left[\frac{(c_a-c_b)\pi}{2\gamma}\right]
 \frac{\sin\left[\pi s_{i}-\frac{(2K+1)\pi^2}{2\gamma}\right]} 
             {\sin(\frac{\pi^2}{2\gamma})};\\
\chi_{1/2}^i&=&\frac{2\chi_{1/2}^0}{N}\times
    \cosh\left[\frac{(c_a+c_b)\pi}{2\gamma}\right]
    \cosh\left[\frac{(c_a-c_b)\pi}{2\gamma}\right]
 \frac{\sin\left[\pi s_{i}-\frac{(2K+1)\pi^2}{2\gamma}\right]} 
     {\cos(\frac{\pi^2}{2\gamma})}; \\  
\chi_{1}^i&=&\frac{2\chi_{1}^0}{N}\times  
    \cosh\left[\frac{(c_a+c_b)\pi}{2\gamma}\right]
    \cosh\left[\frac{(c_a-c_b)\pi}{2\gamma}\right]
 \frac{\sin\left[\pi s_{i}-\frac{(2K+1)\pi^2}{2\gamma}\right]} 
             {\sin(\frac{\pi^2}{2\gamma})},
\end{eqnarray}
therefor, the Kondo temperature can be given by
\begin{eqnarray}
T^k_{1/2}&=&\cosh^{-1}\left[\frac{(c_a+c_b)\pi}{2\gamma}\right]
    \cosh^{-1}\left[\frac{(c_a-c_b)\pi}{2\gamma}\right]
 \frac{\cos(\frac{\pi^2}{2\gamma})}
   {2\sin\left[\pi s_{i}-\frac{(2K+1)\pi^2}{2\gamma}\right]},\\
T^k_{1}&=&\cosh^{-1}\left[\frac{(c_a+c_b)\pi}{2\gamma}\right]
    \cosh^{-1}\left[\frac{(c_a-c_b)\pi}{2\gamma}\right]
 \frac{\sin(\frac{\pi^2}{2\gamma})}
   {2\sin\left[\pi s_{i}-\frac{(2K+1)\pi^2}{2\gamma}\right]},
\end{eqnarray}
where $C_s^0$ and $\chi_s^0$ are the specific heat and magnetic
susceptibility of the infinity size system \cite{bab,resh}, and $K$=0, 1,
2, $\cdots$ determined by
the relation  $ 0< 2s_i\gamma-2K\pi <2\pi. $  
One can check the results eq.(43)-(46) to be
sound based on the fact the $c_i$ and $\chi_i$ proportion to the shift of 
the state density at Fermi surface which is proportional to
$\rho_i(v)/\rho_\infty(v)$ ( see eq.(11)and (12) ), where $\rho_i(v)$ and
$\rho_\infty(v)$ are the density 
of the impurity and the bulk. If we choose $c_a=c_b$ and $s_i=1/2$,
the results of eq.(43), (45) and (47) can recover those in
Ref.\cite{HP}.
In eq.(47) and (48), when $c\rightarrow ic $, the Kondo temperature
satisfy the relation $T_s^k\sim \cos^{-1}[(c_a+c_b)\pi/2\gamma]
\cos^{-1}[(c_a-c_b)\pi/2\gamma]$. The result shows a
crossover from an exponential law to a power law as the phenomenon in
Ref.\cite{lt}. 
 
\section{discussion} \par
In this paper, we obtain the contribution of
the boundary arbitrary spin impurities to the susceptibility, the
specific heat and the Kondo temperature by using the standard method at
low temperature. The model that XXZ chain coupled with boundary impurities
is relevant to the Kondo problem in a Luttinger liquid. And the effect of
the impurities to the bulk can be regarded as a perturbation  to
the bulk. From eq.(43)-(46), we can see that the contribution of
impurities strongly depend on the coupling constants $c_i\;(i=a\,,b)$
and the spin of the impurities $s_i$. Also from these equations, the
Wilson ratio can be obtained as $(C^i_s/C^0_s)/(\xi^i_s/\xi^0_s)=1$ in
this model. This fact can prove that the present model satisfies the
theory of the  Luttinger liquid. Besides the properties discussed in this 
paper, other properties of the present model are also worth
investigating, such as the critical properties of the present model. And
we shall study them in our further papers.

\end{document}